\documentclass[12pt]{iopart}

\usepackage{iopams}

\newtheorem{theorem}{Theorem}[section]

\newtheorem{lemma}[theorem]{Lemma}

\def\IC{{\bf C}}
\def\IR{{\bf R}}

\def\x{{\mathbf x}}
\def\y{{\mathbf y}}

\def\b0{{\bf 0}}
\def\s{{\mathbf{s}}}

\def\cD{{\mathcal D}}

\def\cS{{\mathcal S}}

\def\cH{{\mathcal H}}

\def\cS{{\mathcal S}}

\def\cH{{\mathcal H}}

\def\tr{{\rm trace}\,}

\def\vec{{\rm vec}\,}

\def\[{\left [}
\def\]{\right ]}
\def\({\left (}
\def\){\right )}

\def\dfrac{\displaystyle\frac}

\def\1{{\bf 1}}
\def\>{{\rangle}}
\def\<{{\langle}}

\def\b{{\beta}}

\def\qed{\hfill$\Box$\medskip}

\begin{document}
\openup 1.2\jot

\title[A note on the realignment criterion]
{A note on the realignment criterion}

\author{Chi-Kwong Li$^1$, Yiu-Tung Poon$^2$ and Nung-Sing Sze$^3$}

\address{$^1$ Department of Mathematics, College of William \& Mary,
Williamsburg, VA 23185, USA}

\address{$^2$ Department of Mathematics, Iowa State University,
Ames, IA 50011}

\address{$^3$ Department of Applied Mathematics, The Hong Kong Polytechnic University,
Hung Hom, Hong Kong}

\ead{ckli@math.wm.edu, ytpoon@iastate.edu and raymond.sze@inet.polyu.edu.hk}

\pacs{03.67.-a, 03.67.Mn}

\begin{abstract}
For a quantum state in a bipartite system represented
as a density matrix, researchers used the realignment matrix and functions
on its singular values
to study the separability of the quantum state.
We obtain bounds for elementary symmetric functions of singular values
of realignment matrices. This answers some open problems proposed by Lupo, Aniello, and Scardicchio.
As a consequence, we show that the proposed scheme by these authors for testing
separability would not work if the two subsystems of the bipartite system have the same dimension.
\end{abstract}

\maketitle

\section{Introduction}
\setcounter{equation}{0}

Quantum entanglement was first proposed by Einstein, Podolsky, and Rosen \cite{EPR}
and Schr\"odinger \cite{S35} as a strange phenomenon
of quantum mechanics, criticizing the completeness of the quantum theory.
Nowadays, entanglement is not only regarded as a key for the interpretation of
quantum mechanics or as a mere scientific curiosity,
but also as a resource for various applications, like
quantum cryptography \cite{E91}, quantum
teleportation \cite{BBCJPW}, and quantum computation \cite{RB01}.

\medskip
Suppose quantum states of two quantum systems are
represented by density matrices (positive semidefinite
matrices with trace 1) of sizes $m$ and $n$, respectively.
States of their {\it bipartite composition system} are
represented by  $mn \times mn$ density matrices.
Such a state is {\it separable}  if there are positive
numbers $p_j$ summing up to 1, $m\times m$ density matrices
$\rho^1_j$, and  $n\times n$ density matrices $\rho^2_j$ such that
$$\rho=\sum_{j=1}^k p_j\, \rho^1_j\otimes\rho^2_j.$$

A state is {\it entangled} if it is not separable.
In quantum information science, it is important to determine
the separability of a state. However, the problem of characterizing
separable states is \emph{NP-hard} \cite{Gur03}.
Therefore, researchers focus on finding effective criterion
to determine whether a density matrix is separable or not.

A simple and strong criterion for separability of density matrix is the
{\it computable cross norm or realignment (CCNR) criterion}.
The name CCNR comes from
the fact that this criterion has been discovered in two different forms, namely,
by cross norms \cite{R00,R05} and
by realignment of density matrices \cite{CW03}.

To describe the realignment criterion, let $M_N$ be the set of $N\times N$ complex matrices. $\cD(m,n)$ will denote the set of all $mn\times mn$ density matrices
and $\cD_s(m,n)$ the set of separable density matrices in $\cD(m,n)$.
For any $X=\[x_{ij}\]\in M_n$, let
$$\vec(X) = (x_{11}, x_{12}, \dots, x_{1n},\ x_{21}, x_{22}, \dots, x_{2n},\ \dots,\ x_{n1},x_{n2},\dots, x_{nn}).$$
 If $\rho = \[X_{rs}\]_{1 \le r, s \le m} \in \cD(m,n)$
with $X_{rs} \in M_n$, then
 the realignment of $\rho$  is the $m^2\times n^2$ matrix $\rho^R$ with rows $$\vec(X_{11}), \vec(X_{12}), \dots, \vec(X_{1m}), \vec(X_{21}), \dots, \vec(X_{2m}),
\dots, \vec(X_{m1}), \dots \vec(X_{mm}).$$
For example, if $(m,n) = (2,3)$ and $\rho =\left[\begin{array}{cc} X_{11}&X_{12}\\ X_{21}&X_{22}\end{array}\right ]\in D(2,3)$ with $X_{rs}\in M_3$, then
$$\rho^R = \left [ \begin{array}{c} \vec(X_{11}) \\ \vec(X_{12}) \\ \vec(X_{21}) \\  \vec(X_{22}) \end{array}\right ].$$

The realignment criterion asserts that if $\rho \in \cD_s(m,n)$ then the sum of the
singular values of $\rho^R$ is at most 1. Recall that the {\it singular values}
of an $M\times N$ matrix  $A$ are the nonnegative square roots of the $k = \min\{M,N\}$
largest eigenvalues of the matrix $AA^\dagger$.

For convenience of notation, we assume that $m\le n$ in the following discussion.
For $\rho\in \cD(m,n)$,
let $s_1\ge\cdots\ge s_{m^2}$ be the singular values of $\rho^R$.
The realignment criterion can be stated as
$$s_1+\cdots+s_{m^2}\le 1 \qquad \hbox{ for } \rho\in \cD_s(m,n).$$
In \cite{LAS}, Lupo,   Aniello, and Scardicchio suggest further study of the symmetric functions  on the singular values of $\rho^R$, in order to find conditions beyond the realignment criterion to identify entanglement.

Let
$$\hspace{-1mm}
\begin{array}{rl}
  \cS(m,n) =
  \left\{\(s_1,\dots,s_{m^2}\):\right.&\s_1 \ge \cdots \ge s_{m^2}
\mbox{ are the singular values of }\rho^R,\\ &\left. \mbox{for some }\rho\in \cD(m,n) \right\}\\ &\\
\cS_s(m,n)  =  \left\{\(s_1,\dots,s_{m^2}\):\right.&\s_1 \ge \cdots \ge s_{m^2}
\mbox{ are the singular values of }\rho^R,\\ &\left.
\mbox{for some }\rho\in \cD_s(m,n) \right\}. \end{array}$$
For each $1<\ell\le {m^2}$, define the $\ell$-th elementary symmetric function
$$f_\ell\(s_1,\dots,s_{m^2}\)
=\sum_{1\le i_1<\cdots<i_\ell\le {m^2}}\Pi_{j=1}^\ell s_{i_j}.$$
Following \cite{LAS}, we define for each $1<\ell\le m^2$,
$$\begin{array}{rl}
\tilde B_\ell(m,n)
=&\max \{f_\ell\(\s\):\s \in \cS(m,n),\ \s = \(s_1,\dots,s_{m^2}\) \hbox{ with } \sum_{i=1}^{m^2}s_i\le 1\},\\[3mm]
B_\ell(m,n)=&\max\{f_\ell\(\s\):\s \in \cS_s(m,n)\}.
\end{array}$$
The bounds $\tilde B_\ell(m,n)$ and $ B_\ell(m,n)$
were introduced in \cite{LAS}
using different notations, namely,
$\tilde x_\ell(d,D)$ and $ x_\ell(d,D)$
with $(d,D) =(m^2,n^2)$.

It follows from the definitions  that if $\tilde B_\ell(m,n)> B_\ell(m,n)$, then
there exists an entangled density matrix $\rho$ such that the sum of singular
values of $\rho^R$ is at most $1$ 
but $f_\ell\(s_1,\dots,s_{m^2}\)>B_\ell(m,n)$.
Therefore, the bound $B_\ell(m,n)$ can be used to detect entanglement for which the
realignment criterion fails.
Numerical estimations for these bounds were given for $(m,n)=(2,2)$ and $(2,3)$ in \cite{LAS}.
The numerical results also suggest that $\tilde B_\ell(2,2)= B_\ell(2,2)$ and
$\tilde B_\ell(2,3)> B_\ell(2,3)$. The authors of \cite{LAS} raised the following two open problems in the search for criterion for entanglement beyond the realignment criterion:
\begin{itemize}
\item[(P1)] To determine the actual values of the upper bounds $B_\ell(m,n)$ and $\tilde  B_\ell(m,n)$.
\item[(P2)] To determine if $\tilde B_\ell(m,n)> B_\ell(m,n)$.
\end{itemize}
In this paper, we  study the singular values of $\rho^R$ for a density matrix $\rho$. We refine some  inequalities given in \cite{LAS}.    This leads to an explicit formula for $\tilde B_\ell(m,n)$,
for all $n \ge m$, except for $m^3-m/2<n<m^3$, that gives a partial solution to (P1).
Furthermore, we  show that $\tilde B_\ell(n,n)= B_\ell(n,n)$ for all $n$ and this implies that
the answer to (P2) is negative if $m = n$.

We conclude this section with a reformulation of another simple and strong criterion for separability
in terms of the singular values.
Let $X=\[X_{rs}\]_{1\le r, s\le m}\in \cD(m,n)$ with $X_{rs}\in M_n$.
The {\it partial transpose} of $X$ with respect to the second subsystem is given by $X^{T_2}=\[X_{rs}^t\]_{1\le r, s\le m}$, where $X_{rs}^t$ is the transpose of $X_{rs}$.
The PPT criterion \cite{Per} states that if  $X\in \cD_s(m,n)$, then $X^{T_2}$ is positive semi-definite.
For $m+n\le 5$, PPT criterion is a necessary and sufficient condition for separability \cite{Ho1},
i.e.  $X\in \cD_s(m,n)$ if and only if  $X^{T_2}\in \cD(m,n)$. For $m,n>1$ and $m+n>5$, the PPT criterion
and the CCNR criterion are independent. Note  that for  $X\in \cD(m,n)$,  $X^{T_2}$ is Hermitian. So the  singular values of $X^{T_2}$ are the absolute values  of the eigenvalues of $X^{T_2}$. Since the sum of all eigenvalues of $X^{T_2}$ is equal to $\tr\(X^{T_2}\)=\tr(X)=1$,  $X^{T_2}$ is positive
semi-definite if and only if the sum of the singular values of $X^{T_2}$ is at most $1$, cf. \cite[Corollary 1]{Ho2}.
Thus the PPT criterion shares a similar form with the CCNR criterion.


\section{Main results and their implications}

In this section, we continue to use the notations introduced in Section 1 and
assume that $m\le  n$. We will describe the results and their implications.
The proofs will be given in the next section.

For any density matrix $\rho$,
we obtain the following lower bound for the largest singular value for $\rho^R$,
the realigned matrix of $\rho$.

\begin{lemma} \label{2.1}
Let $\s = (s_1,\dots,s_{m^2})\in \cS(m,n)$. Then $s_1\ge \dfrac{1}{\sqrt{mn}}\,.$
\end{lemma}


Recall that for two vectors $\x,\ \y\in \IR^N$,
$\x$ is {\it  majorized} by $\y$, denotes by $\x\prec \y$,
if for all $1\le k\le N$, the sum of the $k$ largest entries of $\x$ is not larger than that of $\y$,
and  the sum of all entries of $\x$ is equal to that of $\y$.
A function $f: \IR^N \rightarrow \IR$ is {\it Schur concave} if $f(\y) \le f(\x)$
whenever $\x \prec \y$.

Using Lemma \ref{2.1}, we will show that if $n \le m^3$,
then the vector $\s$ in $\cS(m,n)$
always {\it marojize} a vector of the form $(\alpha, \beta, \dots, \beta)$.
One can then apply the theory of majorization and Schur concave functions (see \cite{MO})
to obtain the inequality
$f_{\ell}(\s) \le f_{\ell}(\alpha,\beta, \dots, \beta)$, as shown in Lemma \ref{2.2}.

For $1\le r\le N$, $N\choose r$ will denote the binomial coefficient $\frac{N!}{r!(N-r)!}$.

\begin{lemma} \label{2.2}
Suppose $n\le m^3$ and $\s = (s_1,\dots,s_{m^2})\in  \cS(m,n)$ with $\sum_{i=1}^{m^2}s_i\le 1$.
Let
$$\alpha = \dfrac{1}{\sqrt{mn}} \quad \hbox{ and } \quad
\beta=\dfrac{1-\alpha}{m^2-1} = \frac{\sqrt{mn}-1}{\sqrt{mn}(m^2-1)}.$$
Then
$$(\alpha, \overbrace{\beta, \dots, \beta}^{m^2-1}) \prec \dfrac{1}{\sum_{i=1}^{m^2}s_i}(s_1,\dots,s_{m^2}),$$
and for  $1<\ell\le m^2$,
$$
f_{\ell}(\s)\le f_{\ell}\(\alpha, \beta,\dots,\beta\)\le
\, {{m^2} \choose\ell }\(\frac{1}{m^2}\)^\ell.
$$
Furthermore,
\begin{enumerate}
\item[\rm (a)] $f_{\ell}(\s)  = f_{\ell}\(\alpha, \beta,\dots,\beta\)$ if and only if
$\s=\(\alpha, \beta,\dots,\beta\)$;

\item[\rm (b)] $f_{\ell}\(\alpha, \beta,\dots,\beta\)=
\, {{m^2} \choose\ell }\(\frac{1}{m^2}\)^\ell$ if and only if $n=m^3$.
\end{enumerate}
\end{lemma}

It follows from Lemma \ref{2.2} that
$\tilde B_\ell(m,n) \le \, {{m^2} \choose\ell }\(\frac{1}{m^2}\)^\ell$ for all $m \le n \le m^3$
and the equality holds if and only if $n = m^3$,
which has been shown in  \cite[Proposition 4]{LAS}.
The following result gives an explicit formula for  $\tilde B_\ell(m,n)$
for all $n\ge m$, except for  $m^3-m/2<n<m^3$. This provides a partial solution to problem (P1).

\begin{theorem} \label{2.3}
Suppose  $m\le n\le m^3-m/2$.
Then for  $1<\ell\le m^2$,
$$\tilde B_\ell(m,n)= f_\ell(\alpha,\beta,\dots,\beta),
\quad \mbox{with}\quad
\alpha=\dfrac{1}{\sqrt{mn}} \ \mbox{ and } \ \beta=\dfrac{1-\alpha}{m^2-1}.
$$
If $n\ge m^3$, then $\tilde B_\ell(m,n)= f_\ell(1/m^2,\dots,1/m^2) = \, {{m^2} \choose\ell }\(\frac{1}{m^2}\)^\ell$.

\end{theorem}

\medskip

Theorem \ref{2.3} gives the values of $\tilde B_\ell(m,n)$ for all $n\ge m$, except for
$m^3-m/2<n<m^3$.  In particular, it holds for all  $n$ which is divisible by $m$.
In application, both $n$ and $m$ are powers of $2$. Therefore, $n$ is always divisible by $m$ and $\tilde B_\ell(m,n)$ is given by the above theorem.

When $m = n$, following our proof of Theorem \ref{2.3} in the next section,
one actually gives explicit formulas for $B_\ell(n,n)$ and $\tilde B_\ell(n,n)$.
\begin{theorem} \label{2.4}
For any $n$ and $1\le \ell \le n^2$,
$$B_\ell(n,n)= \tilde B_\ell(n,n)
= f_\ell(\alpha,\beta,\dots,\beta)
\quad\hbox{with}\quad
\alpha=\dfrac{1}{n} \ \mbox{ and } \ \beta=\dfrac{n-1}{n(n^2-1)}.
$$
\end{theorem}
Theorem 2.4 provides partial solutions to both problems (P1) and (P2).
In particular, it gives a negative answer to problem (P2) for the case when $m=n$.
As a result, if $m = n$, the upper bounds of the elementary symmetric functions of
realignment matrices cannot be used to derive new conditions for detecting separability
beyond the realignment criterion.

\section{Proofs}

\bf Proof of Lemma \ref{2.1}. \rm Define $\x=(x_1,\dots,x_{m^2})^t$, $\y=(y_1,\dots,y_{n^2})^t$ by
$$\begin{array}{rl}x_i=&\left\{\begin{array}{ll}
1\quad&\mbox{ if }  i=k(m+1)+1\mbox{ for some }0\le k\le m-1, \\[2mm]
0\quad&\mbox{ otherwise, }
\end{array}\right.
\end{array}$$
and
$$\begin{array}{rl}
y_j=&\left\{\begin{array}{ll}
1\quad&\mbox{ if }  j=k(n+1)+1\mbox{ for some }0\le k\le n-1, \\[2mm]
0\quad&\mbox{ otherwise. }
\end{array}\right.
\end{array}$$
Then  $\dfrac{1}{\sqrt{m}}\, \x$ and $\dfrac{1}{\sqrt{n}}\, \y$ are unit vectors and
$$\dfrac{1}{\sqrt{m}}\, \x^t\, \rho^R\, \dfrac{1}{\sqrt{n}}\, \y
=\dfrac{\tr \rho}{\sqrt{mn}}=\dfrac{1}{\sqrt{mn}}.$$
Because
$$s_1 = \max\left\{ {\bf u}^t \rho^R {\bf v}: {\bf u} \in \IC^{m^2} \hbox{ and }
{\bf v} \in \IC^{n^2} \hbox{ are unit vectors} \right\},$$
we conclude that $s_1\ge \dfrac{1}{\sqrt{mn}}$.
\qed

\medskip\noindent
\bf Proof of Lemma \ref{2.2}. \rm
Note that
$$n \le m^3 \quad \Longleftrightarrow \quad
mn\le m^4 \quad \Longleftrightarrow \quad
\sqrt{mn}-1\le m^2-1
\quad \Longleftrightarrow \quad
\beta\le\alpha\,.$$
Suppose $\s =\(s_1,\dots,s_{m^2}\) \in \cS(m,n)$ with $s = \sum_{i=1}^{m^2}s_i\le 1$.
Let $\tilde \s= (1/s)\, \s$. Then $\tilde s_1\ge s_1\ge \alpha$.
Therefore, $ (1/m^2,\dots,1/m^2)\prec \(\alpha, \beta,\dots,\beta\)
\prec \tilde \s$.
Since $f_{\ell}$ is strictly   concave \cite{MO},  we have
$$f_\ell(\s)\le f_\ell(\tilde\s)\le f_{\ell}\(\alpha, \beta,\dots,\beta\)
\le f_{\ell} (1/m^2,\dots,1/m^2)
=
\, {{m^2} \choose\ell }\(\frac{1}{m^2}\)^\ell, $$
and the equality
$f_{\ell}(\s)  = f_{\ell}\(\alpha, \beta,\dots,\beta\)$
holds if and only if $\s=\(\alpha, \beta,\dots,\beta\)$.
This proves (a). Assertion (b) follows readily from (a).
\qed

\medskip\noindent
\bf Proof of Theorem \ref{2.3}. \rm
We first consider the simpler case when $n \ge m^3$. It suffices to construct
$\rho\in \cD(m,n)$ for which $\rho^R$ has singular values $1/m^2,\dots,1/m^2$.
Suppose $\{E_{1,1},\dots,E_{m,m}\}$ is the standard basis of $m \times m$ matrices.
For $1\le k,\ell\le m$, let
$F_{k,\ell} = (E_{k,\ell} \otimes I_{m^2}) \oplus O_{n-m^3}$.
Then $\rho=\dfrac{1}{m^3}\sum_{k,\,\ell=1}^mE_{k,\ell}\otimes F_{k,\ell}$
is an  $mn \times mn$ density matrix
while $\rho^R$ has singular values $1/m^2,\dots,1/m^2$.

Next, suppose  $m\le n\le m^3-m/2$. By Lemma \ref{2.2},
we have $\tilde B_\ell(m,n)\le f_\ell(\alpha,\beta,\dots,\beta)$
for all $1 \le \ell \le m^2$.
We will construct $\rho\in \cD(m,n)$ for which $\rho^R$ has singular values $\alpha,\beta,\dots,\beta$.
Suppose $n=mq+r$ with $0\le r<m$. For $1\le k,\ell\le m$, let
$F_{k,\ell} = (E_{k,\ell} \otimes I_{q}) \oplus O_{r}$.
Define
$$\rho_1=\sum_{k,\,\ell=1}^mE_{k,\ell}\otimes F_{k,\ell},
\quad
\rho_2 = I_m \otimes \(I_{mq}\oplus O_r\),
\quad\hbox{and}\quad
\rho_3 = I_m \otimes \( O_{mq}\oplus I_{r}\)$$
and
$$\rho =  s_1 \rho_1 + s_2 \rho_2 + s_3 \rho_3
\quad\hbox{with}\quad
s_1 = \frac{\beta}{\sqrt q},
\quad
s_2 = \alpha^2 - \frac{\beta}{m \sqrt q},
\quad\hbox{and}\quad
s_3 = \alpha^2.
$$
Denote $J_{m,n}$ by the $m\times n$ matrix with all entries equal to one.
Then the realigned matrix $\rho^R$ is (under permutation of rows and columns) given by
$$A=\[\begin{array}{c|c|c|c}\overbrace{
s_1 I_{m} + s_2 J_{m,m}\, | \cdots |\,
s_1 I_{m} + s_2 J_{m,m} }^{q-{\rm terms}}& s_3 J_{m,r}&O_{m,(m^2-m)q}&O \\
\hline
&&&\\
O&O&\underbrace{s_1 I_{m^2-m}\, |\cdots |\,  s_1 I_{m^2-m}}_{q-{\rm terms}}&O
\end{array} \].$$
Note that
$$AA^\dagger =\(qs_1^2I_m+(2qs_1s_2+qms_2^2+rs_3^2)J_{m,m}\)\oplus qs_1^2I_{m^2-m}.$$
Since $J_{m,m} $ has only one non-zero eigenvalue $m$,
a matrix of the form $\mu I_m + \nu J_{m,m}$ has
eigenvalues $\mu + m\nu$ and $\mu$ with
multiplicity $1$ and $m-1$, respectively.
As a result $AA^\dagger$ has one eigenvalue equal to
$$
qs_1^2+m(2qs_1s_2+qms_2^2+rs_3^2)= \alpha^4(m^2q+mr) = \alpha^2$$
and $m^2-1$ eigenvalues equal to
$$qs_1^2 = \beta^2.$$
Hence, taking square roots, we see that the matrix $\rho^R$ has the desired singular values $\alpha, \beta, \dots, \beta$.

It remains to show that $\rho$ is a density matrix. Notice that
$$\tr(\rho)
= s_1 (mq) + s_2 (m^2 q) + s_3 (mr)
= \alpha^2 m (m q + r) = 1.$$
Since $\rho_1$, $\rho_2$, and $\rho_3$ are all positive semi-definite
and both $s_1$ and $s_3$ are nonnegative,
$\rho$ is a density matrix if $s_2$ is nonnegative.
Notice that
$$s_2 \ge 0
\quad \Longleftrightarrow \quad
\frac{1}{mn} \ge \frac{\sqrt{mn} - 1}{\sqrt{mn} \sqrt q m(m^2-1)}
\quad \Longleftrightarrow \quad
m^2 - 1 \ge \sqrt \frac{n^2}{q} - \sqrt\frac{n}{mq}.$$
For a fixed $m$, let
$$f(q,r) = \sqrt{\dfrac{(mq+r)^2}{q}}-\sqrt{\dfrac{(mq+r)}{mq}}
\quad\hbox{for}\quad
q \ge 1\hbox{ and }
0 \le r \le m-1.$$
Then
$$\dfrac{\partial f}{\partial q}
=\dfrac{mq-r}{2q^{3/2}}+\dfrac{r}{2q\sqrt{mq(mq+r)}} > 0
\quad\hbox{and}\quad
\dfrac{\partial f}{\partial r}
=\dfrac{1}{\sqrt{q}}-\dfrac{1}{2mq}\sqrt{\dfrac{mq}{mq+r}}>0$$
for all $q \ge 1$ and $0 \le r \le m-1$.
Therefore,
\begin{enumerate}
\item[\rm (a)]
$f(q,r) \le f(m^2-2,m-1)$ for all $1\le q \le m^2-2$ and $r \le m-1$; and

\item[\rm (b)]
$f(m^2-1, r) \le f(m^2-1, m/2)$ for all $r \le m/2$.
\end{enumerate}
So, it suffices to prove that
$${\rm (1)}\quad f(m^2-2,m-1) \le m^2-1
\qquad \hbox{and} \qquad
{\rm (2)}\quad  f(m^2-1, m/2)\le m^2-1.$$
To prove (1), since $m\ge 2$,  we have
$$
m^4(m^2-2)-(m(m^2-2)+m-1)^2
= 2m^4 + 2m^3 -m^2 - 2m-1 >0.$$
It follows that
$\sqrt{\dfrac{(m(m^2-2)+m-1)^2}{m^2-2}} < m^2$
and hence
$$f(m^2-2,m-1)
= \sqrt{\dfrac{(m(m^2-2)+m-1)^2}{m^2-2}}
- \sqrt{\dfrac{(m(m^2-2)+m-1)}{m(m^2-2)}}
\le m^2-1.$$
To prove (2),
since $m^2-1\le \(m-\dfrac{1}{2m}\)^2$, i.e., $\sqrt{m^2-1}\le \(m-\dfrac{1}{2m}\)$,
we have

$$\sqrt{\dfrac{(m(m^2-1)+m/2)^2}{m^2-1}} = m\sqrt{m^2-1}+\dfrac{m}{2\sqrt{m^2-1}}
\le m\( m-\dfrac{1}{2m}\)+\dfrac{m}{2\sqrt{m^2-1}},$$
and
\begin{eqnarray*}
\hspace{-20mm}
\sqrt{\dfrac{(m(m^2-1)+m/2)}{m(m^2-1)}}
&=& \frac{1}{2}\sqrt{1+\dfrac{m^2}{m^2-1}+\dfrac{2m\(m-\frac{1}{2m}\)}{m^2-1}}\\
&\ge& \frac{1}{2}\sqrt{1+\dfrac{m^2}{m^2-1}+\dfrac{2m}{\sqrt{m^2-1}}}
= \frac{1}{2}\(1+\dfrac{m}{\sqrt{m^2-1}}\).
\end{eqnarray*}
Consequently,
\begin{eqnarray*}
\hspace{-18mm}
f(m^2-1,m/2)
&=& \sqrt{\dfrac{(m(m^2-1)+m/2)^2}{m^2-1}}-\sqrt{\dfrac{(m(m^2-1)+m/2)}{m(m^2-1)}}\\[2mm]
&\le&  m\( m-\dfrac{1}{2m}\)+\dfrac{m}{2\sqrt{m^2-1}} -\dfrac{1}{2}\(1+\dfrac{m}{\sqrt{m^2-1}}\)
= m^2-1 .
\end{eqnarray*}
\vskip -.2in \qed

\medskip\noindent{\bf Remark}\ \
 The smallest values of $m,\ n$ which do not satisfy the conditions in Theorem \ref{2.3} are
$m=3$ and $n=26$. For these values, the proof in Theorem \ref{2.3} does not work because $s_2<0$. In this case, the question about the exact value of $\tilde B_\ell(m,n)$ is still open.

\medskip\noindent
\bf Proof of Theorem \ref{2.4}.
\rm Suppose $m=n$. Then the matrix $\rho$ constructed in the proof of
Theorem \ref{2.3} has the form
$$\rho = \dfrac{1}{n(n+1)}\(I_{n^2}+xx^t \)$$
where
$$x_i =
\left\{\begin{array}{ll}
1\quad&\mbox{if }  i=k(n+1)+1\mbox{ for some }0\le k\le n-1, \\[2mm]
0\quad&\mbox{otherwise. }
\end{array}\right.
$$
It follows from \cite{PR} that $\rho$ is separable. \qed

\section{Conclusion}

The main goal of  this paper  is to investigate the open problems (P1) and (P2) proposed in \cite{LAS} in the search for a new criterion for separability. We study the singular values of the realignment of density matrices and obtain
new bounds on the elementary symmetric functions. The results are
applied to find   explicit formulas for $\tilde B_\ell(m,n)$, for all $n\ge m$, except $m^3-m/2< n< m^3$  and $ B_\ell(n,n)$. This provides a partial answer
to the open problem (P1).
Furthermore, we show that $\tilde B_\ell(n,n) = B_\ell(n,n)$ for all $n$ so that
one cannot use $\tilde B_\ell(m,n)$ to differentiate separable matrices from
density matrices whose realignment matrix has trace norm at most $1$ when $m = n$.
This gives a negative answer to problem (P2) when $m = n$.
For $m\ne n$,  numerical results in \cite{LAS} suggested
that $\tilde B_\ell(m,n) >  B_\ell(m,n)$. If this strict inequality holds, then we would have a new criterion for separability. Our explicit formula for $ \tilde B_\ell(m,n)$ will be useful in this study.


\section*{Acknowledgments}
The research was done when Li was a 2011 Fulbright Fellow at the
Hong Kong University of Science and Technology.
He is a Shanxi Hundred Talent Scholar of the Taiyuan University
of Science and Technology, and
is an honorary professor of the University of Hong Kong,
and  the Shanghai University.
Research of the first two authors were supported in part by USA NSF.
Research of the first and the third authors were supported in part by a HK RGC grant.

\end{document}